\documentclass[sigconf]{acmart}

\AtBeginDocument{%
	}
		
\acmConference[ICSE 2024]{46th International Conference on Software Engineering}{April 2024}{Lisbon, Portugal}

\setcopyright{acmcopyright}

\ccsdesc[500]{Security and privacy~Software security engineering}
\ccsdesc[500]{Computing methodologies~Knowledge representation and reasoning}
\ccsdesc[500]{Security and privacy~Software security engineering}

\usepackage[graphicx]{realboxes}
\usepackage{color}
\usepackage{tcolorbox}
\usepackage{xcolor}
\usepackage{bm}
\usepackage{multirow}

\begin{document}
\copyrightyear{2024}
\acmYear{2024}
\setcopyright{acmlicensed}\acmConference[ICSE '24]{2024 IEEE/ACM 46th International Conference on Software Engineering}{April 14--20, 2024}{Lisbon, Portugal}
\acmBooktitle{2024 IEEE/ACM 46th International Conference on Software Engineering (ICSE '24), April 14--20, 2024, Lisbon, Portugal}
\acmDOI{10.1145/3597503.3639173}
\acmISBN{979-8-4007-0217-4/24/04}

\title{Improving Smart Contract Security with Contrastive Learning-based Vulnerability Detection}

\author{Yizhou Chen}
\affiliation{%
  \institution{Key Lab of HCST (PKU), MOE; \\ School of Computer Science, Peking University}
  \city{Beijing}
  \country{China}}
\email{yizhouchen@stu.pku.edu.cn}

\author{Zeyu Sun}
\authornote{Zeyu Sun is the corresponding author.\\HCST: High Confidence Software Technologies.}
\affiliation{%
  \institution{Science \& Technology on Integrated Information System Laboratory, Institute of Software, Chinese Academy of Sciences}
  \city{Beijing}
  \country{China}}
\email{szy_@pku.edu.cn}

\author{Zhihao Gong}
\affiliation{%
  \institution{Key Lab of HCST (PKU), MOE; \\ School of Computer Science, Peking University}
  \city{Beijing}
  \country{China}}
\email{zhihaogong@stu.pku.edu.cn}

\author{Dan Hao}
\affiliation{%
  \institution{Key Lab of HCST (PKU), MOE; \\ School of Computer Science, Peking University}
  \city{Beijing}
  \country{China}}
\email{haodan@pku.edu.cn}

\begin{abstract}
	Currently, smart contract vulnerabilities (SCVs) have emerged as a major factor threatening the transaction security of blockchain. Existing state-of-the-art methods rely on deep learning to mitigate this threat. They treat each input contract as an independent entity and feed it into a deep learning model to learn vulnerability patterns by fitting vulnerability labels. It is a pity that they disregard the correlation between contracts, failing to consider the commonalities between contracts of the same type and the differences among contracts of different types. As a result, the performance of these methods falls short of the desired level. 
	
	To tackle this problem, we propose a novel Contrastive Learning Enhanced Automated Recognition Approach for Smart Contract Vulnerabilities, named Clear. In particular, Clear employs a contrastive learning (CL) model to capture the fine-grained correlation information among contracts and generates correlation labels based on the relationships between contracts to guide the training process of the CL model. Finally, it combines the correlation and the semantic information of the contract to detect SCVs. Through an empirical evaluation of a large-scale real-world dataset of over 40K smart contracts and compare 13 state-of-the-art baseline methods. We show that Clear achieves (1) optimal performance over all baseline methods; (2) 9.73\%-39.99\% higher F1-score than existing deep learning methods.
\end{abstract}
\keywords{Smart contract, Vulnerability detection, Deep learning, Contrastive learning}

\maketitle
\section{Introduction}
\label{sec1}
In contemporary times, transactions based on blockchain systems and smart contracts are becoming increasingly popular in both personal and commercial settings \cite{mariano2020demystifying,singh2020blockchain}. Yet, this increased reliance on blockchain systems has made them a tempting target for cybercriminals seeking to exploit software vulnerabilities for illegal financial gain ~\cite{chen2022blockchain,ren2021making}. With the exponential growth of the virtual currency market, Smart Contract Vulnerabilities (SCVs) have become a major risk threatening secure transactions on the blockchain. The potential exploitation of these vulnerabilities by malicious actors could result in the compromise of virtual assets, putting users at risk of significant financial losses ~\cite{praitheeshan2019security,wan2021smart}. In 2016, the Decentralized Autonomous Organization on Ethereum was attacked, and the attacker exploited an SCV to steal approximately \$50 million worth of ether \cite{choi2021smartian}. Moreover, in 2018, the decentralized exchange Bancor suffered an SCV, resulting in the theft of roughly \$23.5 million worth of cryptocurrencies. The above emergencies reveal that Smart Contract Vulnerability Detection (SCVD) has become an urgent task.

\begin{figure*}
	\centering
	\includegraphics[width=0.8\textwidth]{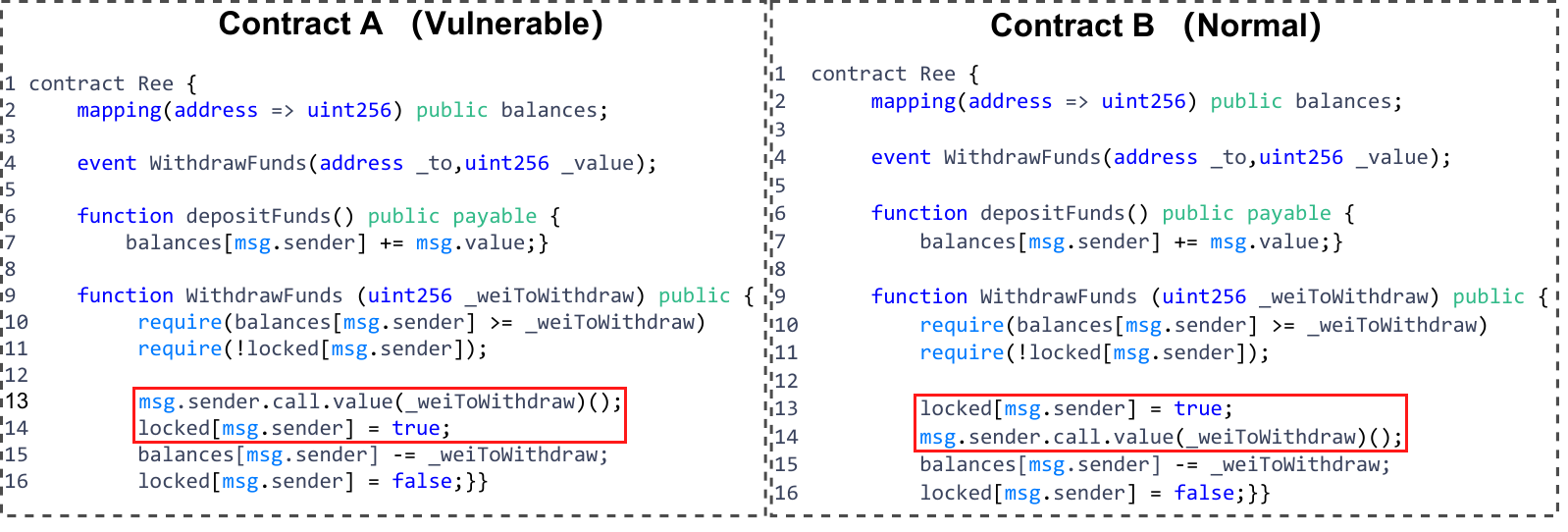}
	\caption{An example of smart contracts.}
	\label{fig:example_sc}
\end{figure*}

To detect SCVs, numerous researchers proposed effective methods, which are broadly divided into two categories. 
The first line of work ~\cite{feist2019slither,luu2016making,mueller2017framework,nguyen2020sfuzz,tikhomirov2018smartcheck,torres2018osiris,tsankov2018securify} is rule-based techniques, which identify SCVs through predefined rules or manually-defined patterns on the smart contract code and its execution. 
However, the vulnerabilities that exist in smart contracts are diverse, which can make predefined patterns insufficient in covering all possible vulnerability types. As a result, these methods may produce false positives or false negatives, which undermine their effectiveness in detecting vulnerabilities accurately ~\cite{qian2023cross}. Moreover, developing these patterns is a time-consuming and error-prone process that relies heavily on manual work. Therefore, the researchers recognize the need to explore alternative approaches that can help reduce the cost and improve the accuracy of SCVD.

Another line of work ~\cite{qian2023cross,kipf2016semi,liu2021smart,zhuang2021smart} utilizes deep learning methods to automatically detect SCVs, resulting in impressive performance gains. These methods use neural networks to learn the vulnerability patterns and detect SCVs.
The commonality of these methods is that they treat each input contract as an isolated entity labeled with whether it is vulnerable. Indeed, a contract usually contains many lines of code, but only a few are relevant to SCVs. Some fine-grained information can hardly be learned by the existing methods, but can be caught by the difference between the vulnerable and non-vulnerable contracts. In other words, these deep learning methods have achieved limited performance because they ignore the correlation between contracts, including the difference between vulnerable and non-vulnerable contracts, as well as the commonalities between vulnerable contracts. 

To solve the problem, we propose a novel Contrastive Learning Enhanced Automated Recognition approach for SCVs, called Clear. Clear introduces the contract correlation into the field of SCVD and leverages the contrastive learning (CL) model to learn pair-wise comparisons of smart contracts and find their correlations. 
In addition, we guide the training process of the CL model by reusing existing vulnerability labels to generate correlation labels. These efforts are used to improve the performance of SCVD.
To outline briefly, Clear samples pairs of contracts from the dataset and generate a correlation label for the contract pairs.
Then, a CL model is constructed, which consists of a contextual augmentation module, a Transformer module, and a contrastive loss function, to learn the fine-grained correlation information between pairs of contracts by fitting correlation labels. Finally, we fine-tune the Transformer module and fit vulnerability labels to enhance the performance of vulnerability detection.

Our proposed method has been rigorously evaluated on the largest established dataset on SCVD, which consists of over 40K real-world smart contracts, by comparing it against 13 state-of-the-art SCVD methods.
The quantitative experimental results show the proposed Clear outperforms all the state-of-the-art methods across all metrics. In particular, Clear achieves significant improvement over even the best-performing method DMT ~\cite{qian2023cross}: precision increased from 87.28\% to 93.64\% (by 7.29\%), recall improved from 85.13\% to 95.44\% (by 12.11\%), and F1-score elevated from 86.14\% to 94.52\% (by 9.73\%) on three types of SCVs, averagely. Besides, our ablation study shows that Clear achieves outperformance by clustering vulnerability contracts in the feature space and separating them from non-vulnerability contracts. Moreover, we experimentally demonstrate that the proposed CL model enhances RNN-based models (i.e., RNN, LSTM, GRU) and boosts their performance by 40.51\%-50.94\% in terms of the F1-score compared to the original model.
In summary, this paper makes the following contributions.

\begin{itemize}
	\item \textbf{A contrastive-learning-based vulnerability detection technique Clear}, which utilizes fine-grained correlation information among smart contracts to improve the performance of SCVD.
	\item \textbf{An extensive experiment} on a large-scale dataset of over 40,000 smart contracts comparing against 13 state-of-art SCV methods, which shows the effectiveness of Clear.
	\item \textbf{A reproducible package} available at \url{https://github.com/chenpp1881/Clear}.
\end{itemize} 

\begin{figure*}
	\centering
	\includegraphics[width=\textwidth]{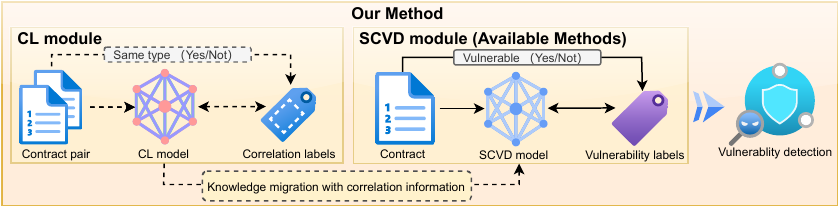}
	\caption{Architecture of our method and the available methods.}
	\label{fig:idea}
\end{figure*}

\section{Motivating Examples}
Smart contract development is a relatively unfamiliar field, and numerous developers may lack an in-depth understanding of smart contract security. They may focus on the functionality and business logic of the contract while ignoring potential security risks, thus introducing vulnerabilities in the code-writing process. Moreover, developers often unknowingly make small mistakes that can result in vulnerabilities. 
As a result, contracts that exhibit vulnerabilities can closely resemble those that do not. 
Figure \ref{fig:example_sc} shows an example where both contracts have identical functionality, with the only distinction residing in the order of statements within the ``WithdrawFunds'' function (lines 13-16). In \textit{Contract B}, the account is initially locked (line 13), followed by the updating and transfer of the account balance (lines 14-16). Conversely, in \textit{Contract A}, the transfer balance is executed (line 13) before the account is locked (line 14). This minor difference in \textit{Contract A} introduces a vulnerability. 
In practical development scenarios, the logic of smart contracts is often highly complex, leading to the frequent occurrence of the aforementioned situation. Unfortunately, current SCVD methods treat smart contracts as isolated entities and rely on deep learning models to learn vulnerability features or patterns to identify SCVs. 
These methods may overlook vulnerabilities triggered by subtle faults. Specifically, the limitations of existing SCVD methods stem from their architecture, which requires deep learning models to independently explore and learn semantic information from the entire contract code, guided by vulnerability labels, to identify possible vulnerability patterns. This architecture lacks sufficient detail for deep learning models to adequately comprehend the fundamental nature of vulnerabilities. The challenges faced by deep learning models in accurately capturing correlation information among contracts under this architecture encompass fine-grained differences between vulnerable and non-vulnerable contracts, as well as commonalities among vulnerable contracts. Undoubtedly, correlation information plays a crucial role in effectively identifying SCVs.

However, until now, the impact of contract correlations on SCVD is still unexplored in existing studies. The CL model provides important inspiration for our work. The CL model was originally developed as an unsupervised learning technique in the field of computer vision to learn representations by uncovering the underlying similarities and dissimilarities among samples ~\cite{khosla2020supervised,chuang2020debiased,xiao2020should,chen2020simple}. Therefore, to address the aforementioned issues, we extend it to the SCVD tasks of supervised learning by adapting the methodology used for constructing sample pairs and correlation labels. To be specific, as shown in Figure \ref{fig:idea}, our method diverges from existing SCVD methods, as we initially leverage the CL model to learn the correlations among contracts. Subsequently, we migrate the correlation knowledge to the SCVD model, integrating it with the vulnerability features of the contracts to accurately detect and identify SCVs. It is worth emphasizing that our method is specifically designed to target common and subtle faults in SCVs. If similar vulnerability characteristics exist in the software of other domains, our method may also be adapted to detect them.

\section{Methodology}

\subsection{Method Overview}
\label{3-1}
In migrating the CL model to the SCVD domain, we are faced with three primary issues. (1) It is essential to determine an appropriate label that can guide the CL model in learning effective correlation information. (2) The neural network for contextual semantic representation of smart contracts needs to be refined and designed to improve the generalization of the CL model. (3) Leveraging both correlation information and vulnerability features to enhance the performance of SCVD.

\begin{figure*}
	\centering
	\includegraphics[width=1\textwidth]{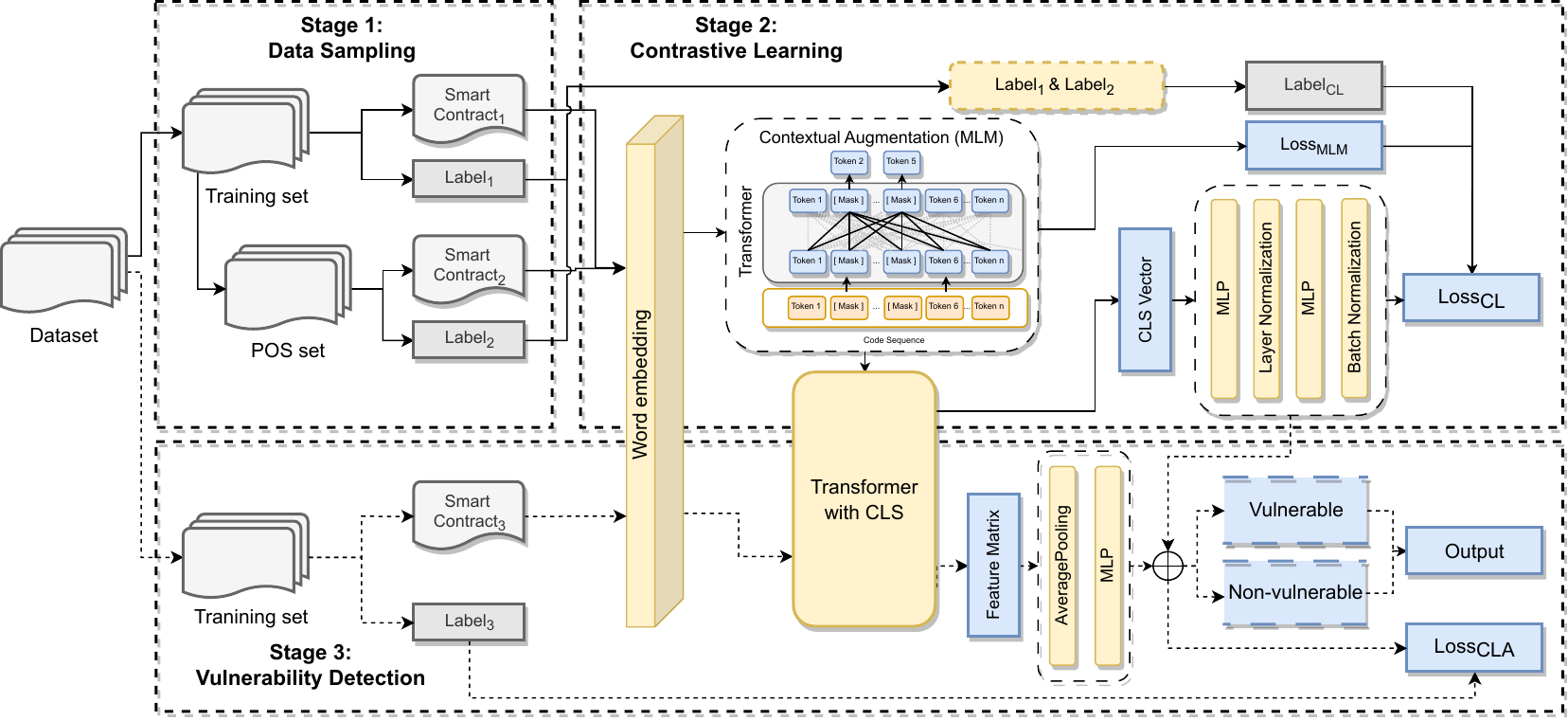}
	\caption{Overview of Clear, which encompasses both the CL process, depicted by solid lines indicating the data flow, and the subsequent vulnerability detection process, represented by dotted arrows indicating the data flow.}
	\label{overview}
\end{figure*}

Therefore, we present a novel approach for SCV automated detection, named Clear, as illustrated in Figure \ref{overview}. Clear sequentially tackles the aforementioned issues through three steps:

\begin{itemize}
	\item[1.] \textbf{Data Sampling:} contract pairs are sampled from the dataset to serve as input for the CL model. A correlation label is assigned to each pair, indicating their relationship and guiding the training process of the CL model.
	\item[2.] \textbf{Contrastive Learning:} we devise a CL model that incorporates a contextual augmentation module and a Transformer-based feature learning module. By computing contrastive loss, the model learns to capture correlations between contract pairs that align with the provided labels.
	\item[3.] \textbf{Vulnerability Detection:} we fine-tune the Transformer model from Stage 2 and combine the outputs of the CL model to recognize SCVs by a fully connected neural network.
\end{itemize}

\subsection{Data Sampling} \label{3-2}
For the CL module, we incorporate correlation labels to guide the training process. These labels are constructed based on the relationships between sampled contracts. Therefore, employing a suitable sampling strategy is crucial as it can greatly enhance the performance by ensuring the utilization of high-quality sample pairs for training, while minimizing the introduction of bias into the learning process. Motivated by this, our sampling strategy is as follows:

There are three types of relationships for contract pairs, i.e. ``\textbf{V}-\textbf{V}'', ``\textbf{N}-\textbf{N}'', and ``\textbf{V}-\textbf{N}'', where \textbf{V} and \textbf{N} denote \textbf{V}ulnerable and \textbf{N}on-vulnerable contracts, respectively. Our intuition is that the relationships of ``\textbf{V}-\textbf{V}'' and ``\textbf{V}-\textbf{N}'' are more important, because we would like to discover the commonality in ``\textbf{V}-\textbf{V}'' and differences in ``\textbf{V}-\textbf{N}'' by CL. In contrast, the ``\textbf{N}-\textbf{N}'' is not substantially helpful in identifying SCV. Therefore, our sampling strategy is to extract all vulnerable contracts from the original dataset and create a new set called the POS set. Then, for each contract in the original dataset, we randomly select a contract from the POS set to form a pair of contracts as input for the CL model. It should be noted that this sampling strategy does not have ``\textbf{N}-\textbf{N}'' relationship. Finally, the correlation labels $L_{CL}$ of the contract pairs are constructed to guide the training of the CL model. The rule is as follows:

\begin{equation}
	L_{CL} =
	\begin{cases}
		1, & \text{If ``\textbf{V}-\textbf{V}''} \\
		0, & \text{If ``\textbf{V}-\textbf{N}''}
	\end{cases}.
\end{equation}

\subsection{Contrastive Learning} \label{3-3}
To better model correlations, in the CL stage, we first apply contextual augmentation to the input contract, then use a Transformer to learn contract features, and finally employ the contrastive loss function to optimize the Transformer model. It should be noted that both contracts in a contract pair undergo the same encoding process and share identical model parameters. Therefore, for simplicity and clarity, we only demonstrate the encoding process for a single contract in the following subsections.

\subsubsection{Contextual Augmentation}
To enhance the comprehension of semantic and structural features in the contract code, as well as to facilitate understanding of contract correlation, we incorporate a contextual augmentation module, i.e., masked language model (MLM), at the beginning of the CL stage. The core of MLM is a self-supervised Transformer model. In simple terms, MLM involves randomly masking certain tokens in the input data. The Transformer model is then tasked with predicting the masked tokens based on the surrounding context. 
Predicting the masked tokens prompts the model to discern meaningful patterns, relationships, and dependencies within the contract code, facilitating a more comprehensive understanding of its underlying structure and semantics. This approach helps the model capture important features and contextual information, which can prove advantageous for downstream prediction tasks.

To be specific, given a code sequence, we randomly select 30\% of the tokens to be replaced with a special [Mask] token, and keep the remaining 70\% unchanged. Then, the Transformer is utilized to predict the original token that corresponds to the masked token. The loss function of an MLM can be represented using the cross-entropy loss function as follows:
\begin{equation}
	\text{Loss}_{MLM} = -\sum_{j \in T} y_{j}\log{\hat{y}_{j}},
\end{equation}
where $T$ denotes a set of the index of masked tokens, $y_{j}$ is the ground truth label of the $j$-th masked token, and $\hat{y}_{j}$ is the predicted probability of the $j$-th masked token being the ground truth label. Specifically, in an MLM, only the tokens that are replaced with the special token [MASK] are used to calculate the loss.

\subsubsection{Feature Learning}
\label{subsubsec:transformer}
Our feature learning module also follows a standard Transformer process with a minor modification. Given that the CL module necessitates the computation of distances in the complete sequence representation of two samples, we incorporate CLS vectors as a way to extract the entire semantic information of code samples. This idea is inspired by previous research ~\cite{feng2020codebert}. Incidentally, the introduction of CLS vectors in the CL stage can enhance model efficiency by eliminating additional processes, such as sequence modeling and data alignment, solely contrasting two CLS vectors. The $\bm{CLS} \in \mathbb{R}^{1 \times k}$ vector serves as an additional input token and is generated in the following manner:
\begin{equation}
	\bm{CLS} = \frac{1}{\sqrt{n}} \sum_{i=1}^{n} X'_i,
\end{equation}
where $X' \in \mathbb{R}^{n \times k}$ is the output $X$ of MLM, $k$ is the word embedding dimension. Then, the position encoding $PE$ is employed to furnish token-level positional information for $X$. The specific process is as follows: 
\begin{equation}
	PE_{(pos, 2l)} = \sin \frac{pos}{10000^{2l/k}},
\end{equation}
\begin{equation}
	PE_{(pos, 2l+1)} = \cos \frac{pos}{10000^{2l/k}},
\end{equation}
where $pos$ is the position identifier that records the position information of the token in the sequence, and $l$ denotes the dimension index.

Subsequently, together with $CLS$ and $X'$, it serves as input for the multi-head attention mechanism (MHAM). The mathematical process can be represented as:

\begin{equation}
	\bm{CLS'} , F = \text{MHAM}( \bm{CLS}  \oplus (X' + PE)),
\end{equation}
where $\oplus$ and $+$ denote concatenation and element-wise addition.

Whereafter, the $\bm{CLS'}$ vectors capture global semantic information about the contract and serve as representative summaries of its overall content. They are utilized in the CL stage to establish correlations between instances of contracts. Conversely, the feature vectors $F \in \mathbb{R}^{n \times k}$, which comprise the encoded representations of each token and its contextual dependencies, are utilized in the vulnerability detection stage.

\subsubsection{Contrastive Loss}
We then apply two linear transformations - L2 normalization and batch normalization - to process the $\bm{CLS'}$. Specifically, L2 normalization promotes a more balanced distribution of the vector's elements, preventing any single feature from dominating the learning process. Batch normalization improves model convergence and stability by reducing internal covariate shift, which is the variation in activation distribution across different layers during training. Their mathematical process is as follows:
\begin{equation}
	\label{ClOutput}
	\bm{v} = \text{BatchNorm}(W_2 \cdot \text{LayerNorm}(W_1 \cdot \bm{CLS'})),
\end{equation}
where $W_1$ and $W_2$ are weights of the linear transformation and $\bm{v}$ is the ultimate global vector representation of a contract in the CL stage.

In this way, the contract pair yields a pair of vector representations $[\bm{v}_a,\bm{v}_b]$. Then, we compute the contrastive loss $Loss_{CL}$ with the correlation label $L_{CL}^{ab}$. The contrastive loss is formulated as follows:

\begin{equation}
	\begin{aligned}
		\text{Loss}_{CL}(\bm{v}_a, \bm{v}_b, L_{CL}^{ab}) & = L_{CL}^{ab} \cdot \text{sim}(\bm{v}_a, \bm{v}_b)^2 \\
		& +  (1 - L_{CL}^{ab}) \cdot \max(0, M - \text{sim}(\bm{v}_a, \bm{v}_b))^2, \\
	\end{aligned}
\end{equation}
where $\text{sim}(\cdot)$ represents the euclidean distance of the two vectors, and $M$ is the margin that determines the threshold for dissimilarity.

Finally, the contrastive loss, denoted as $\text{Loss}_{CL}$, and the MLM loss, denoted as $\text{Loss}_{MLM}$, are combined to optimize the model. The overall loss function can be expressed as follows:
\[
\text{Total Loss} = \lambda_{CL} \cdot \text{Loss}_{CL} + \lambda_{MLM} \cdot \text{Loss}_{MLM}.
\]
In this equation, $\lambda_{CL}$ and $\lambda_{MLM}$ are hyperparameters that control the relative importance of the contrastive loss and MLM loss, respectively.

\subsection{Vulnerability Detection} \label{3-4}
During the vulnerability detection stage, we focus on fine-tuning the Transformer encoder mentioned earlier to accurately detect SCV. 
After obtaining the feature representation $F$ of the contracts, we combine $F$ and correlation features $\bm{v}$ to detect SCV using a fully connected neural network. The process can be represented as follows:
\begin{equation}
	\hat{L} = \sigma(W_3 \cdot (\text{AvgPooling}(F) \oplus \bm{v}) + \bm{b}),
\end{equation}
where $\sigma$ is the sigmoid activation function, $W_3$ is the weight matrix and $b$ is the bias vector, $\oplus$ indicates concatenation operation. The output $\hat{L}$ is the predicted probability (vulnerable or non-vulnerable) and calculates the loss with the real vulnerability label $\hat{y} \in (0,1)$ by the cross entropy loss $\text{Loss}_{CLA}$.
\begin{equation}
	\text{Loss}_{CLA}(\hat{L}, \hat{y})=-  \left( \hat{y} \log \left(\hat{L}\right) + \left(1 - \hat y\right) \log \left(1 - \hat{L}\right) \right).
\end{equation}

\begin{table*}[]
	\caption{The performance evaluation of our method is compared with 13 baseline models involving baselines in terms of Recall (R), Precision (P) and F1-score (F). ``n/a'' denotes not applicable.}
	\resizebox{.9\linewidth}{!}{
		\renewcommand{\arraystretch}{1.1}
		\begin{tabular}{l|l|rrr|rrr|rrr|rrr}
			\toprule
			\multirow{4}{*}{\textbf{Line \#}}&\multirow{4}{*}{\textbf{Methods}} & \multicolumn{3}{c|}{\multirow{2}{*}{\textbf{RE}}} & \multicolumn{3}{c|}{\multirow{2}{*}{\textbf{TD}}} & \multicolumn{3}{c|}{\multirow{2}{*}{\textbf{IO}}} & \multicolumn{3}{c}{\multirow{2}{*}{\textbf{Average}}} \\
			&& \multicolumn{3}{c|}{} & \multicolumn{3}{c|}{} & \multicolumn{3}{c|}{} & \multicolumn{3}{c}{} \\ \cline{3-14} 
			&& \multirow{2}{*}{\textbf{R(\%)}} & \multirow{2}{*}{\textbf{P(\%)}} & \multirow{2}{*}{\textbf{F(\%)}} & \multirow{2}{*}{\textbf{R(\%)}} & \multirow{2}{*}{\textbf{P(\%)}} & \multirow{2}{*}{\textbf{F(\%)}} & \multirow{2}{*}{\textbf{R(\%)}} & \multirow{2}{*}{\textbf{P(\%)}} & \multirow{2}{*}{\textbf{F(\%)}} & \multirow{2}{*}{\textbf{R(\%)}} & \multirow{2}{*}{\textbf{P(\%)}} & \multirow{2}{*}{\textbf{F(\%)}} \\
			&&  &  &  &  &  &  &  &  &  &  &  &  \\ \midrule
			1&\textbf{sFuzz} & 14.95 & 10.88 & 12.59 & 27.01 & 23.15 & 24.93 & 47.22 & 58.62 & 52.31 & 29.73 & 30.88 & 29.94 \\
			2&\textbf{Smartcheck} & 16.34 & 45.71 & 24.07 & 79.34 & 47.89 & 59.73 & 56.21 & 45.56 & 50.33 & 50.63 & 46.39 & 44.71 \\
			3&\textbf{Osiris} & 63.88 & 40.94 & 49.90 & 55.42 & 59.26 & 57.28 & n/a & n/a & n/a & n/a & n/a & n/a \\
			4&\textbf{Oyente} & 63.02 & 46.56 & 53.55 & 59.97 & 61.04 & 59.47 & n/a & n/a & n/a & n/a & n/a & n/a \\
			5&\textbf{Mythril} & 75.51 & 42.86 & 54.68 & 49.80 & 57.50 & 53.37 & 62.07 & 72.30 & 66.80 & 62.46 & 57.55 & 58.28 \\
			6&\textbf{Securify} & 73.06 & 68.40 & 70.41 & n/a & n/a & n/a & n/a & n/a & n/a & n/a & n/a & n/a \\
			7&\textbf{Slither} & 73.50 & 74.44 & 73.97 & 67.17 & 69.27 & 68.20 & 52.27 & 70.12 & 59.89 & 64.31 & 71.28 & 67.35 \\ \midrule
			8&\textbf{GCN} & 73.18 & 74.47 & 73.82 & 77.55 & 74.93 & 76.22 & 69.74 & 69.01 & 69.37 & 73.49 & 72.80 & 73.14 \\
			9&\textbf{TMP} & 75.30 & 76.04 & 75.67 & 76.09 & 78.68 & 77.36 & 70.37 & 68.18 & 69.26 & 73.92 & 74.30 & 74.10 \\
			10&\textbf{AME} & 78.45 & 79.62 & 79.03 & 80.26 & 81.42 & 80.84 & 69.40 & 70.25 & 69.82 & 76.04 & 77.10 & 76.56 \\
			11&\textbf{SMS} & 77.48 & 79.46 & 78.46 & 91.09 & 89.15 & 90.11 & 73.69 & 76.97 & 75.29 & 80.75 & 81.86 & 81.29 \\
			12&\textbf{DMT} & 81.06 & 83.62 & 82.32 & 96.39 & 93.60 & 94.97 & 77.93 & 84.61 & 81.13 & 85.13 & 87.28 & 86.14 \\ \midrule
			13&\textbf{LineVul} & 73.01 & 85.19 & 78.63 & 67.46 & 89.47 & 76.92 & 74.20 & 74.10 & 74.15 & 71.56 & 82.92 & 76.57 \\ \midrule
			14&\textbf{Clear} & \textbf{96.43} & \textbf{96.81} & \textbf{96.62} & \textbf{98.41} & \textbf{94.30} & \textbf{96.31} & \textbf{91.48} & \textbf{89.81} & \textbf{90.64} & \textbf{95.44} & \textbf{93.64} & \textbf{94.52} \\ \bottomrule
	\end{tabular}}
	\label{tab:rq1}
\end{table*}

\section{Experimental Settings}


We conduct comprehensive evaluations of our proposed framework to address the following Research Questions (RQ):

\noindent
\textit{RQ1: How does our method Clear perform compared against 13 state-of-the-art SCVD techniques?} 

\noindent
\textit{RQ2: How do the different modules affect the performance of the proposed approach?} 

\noindent
\textit{RQ3: Does our proposed CL module enhance the performance of other deep learning models besides Transformer?}

\subsection{Dataset} \label{sec:dataset}
We select the largest publicly available vulnerability dataset for smart contracts~\cite{qian2023cross}, which consists of 40K real-world smart contracts. The dataset is carefully labeled with distinct types of SCVs. Among the 40K contracts, 4290 contracts were identified to contain vulnerabilities: 680 contracts were identified to possess reentrancy vulnerabilities (RE), 2242 contracts exhibited timestamp dependency vulnerabilities (TD), and 1368 contracts were found to have integer overflow/underflow vulnerabilities (IO). This dataset randomly assigns 80\% contracts as the training set and the remaining 20\% as the test set, and in our evaluation, we use the same split sets.

These vulnerabilities are widely studied in prior work ~\cite{qian2023cross}, because a large portion of the financial loss in Ethereum is attributed to these vulnerabilities~\cite{tikhomirov2018smartcheck,luu2016making,mueller2017framework,tsankov2018securify,ibing2015fixed} and existing research has demonstrated that these vulnerabilities are prevalent in Ethereum smart contracts. In particular, \textbf{RE} occurs when a contract allows an external attacker to re-enter a function before the previous execution completes, leading to unexpected and unauthorized behavior \cite{grech2018madmax}. \textbf{TD} arises from the reliance on block timestamps for critical decisions within smart contracts. Attackers can manipulate timestamps to their advantage, compromising contract logic and enabling activities such as front-running or denial-of-service attacks \cite{luu2016making}. \textbf{IO} occurs when arithmetic operations on integers exceed the maximum or minimum representable values \cite{kalra2018zeus}. These vulnerabilities can result in incorrect calculations, allowing attackers to manipulate values, bypass security checks, and cause financial harm. 

\subsection{Baselines}  \label{sec:baselines}
In our evaluation, we first select a set of baseline methods specifically designed for SCVD. These baselines represent state-of-the-art approaches in the SCVD field. They can be broadly classified into two categories: rule-based techniques and neural network-based techniques.

Rule-based techniques, as the category of baselines, rely on rule-based heuristics to identify SCVs, including sFuzz ~\cite{nguyen2020sfuzz}, Smartcheck ~\cite{tikhomirov2018smartcheck}, Osiris ~\cite{torres2018osiris}, Oyente ~\cite{luu2016making}, Mythril ~\cite{mueller2017framework}, Securify ~\cite{tsankov2018securify}, and Slither ~\cite{feist2019slither}.

Neural network-based methods, on the other hand, leverage deep learning techniques to identify SCVs. Here, we include five state-of-art neural-learning-based SCVD methods, including GCN ~\cite{kipf2016semi}, TMP ~\cite{zhuang2021smart}, AME ~\cite{liu2021smart}, SMS ~\cite{qian2023cross}, and DMT ~\cite{qian2023cross}.

General vulnerability detection methods~\cite{fu2022linevul,li2021vulnerability,zhou2019devign} exist, but none of them yields satisfactory results in the field of SCVD. Therefore, in this paper, we have chosen only the best-performing one, namely LineVul ~\cite{fu2022linevul}, as the representative of general methods. 

The details of all baselines are in Section~\ref{RelatedWork}.

\subsection{Metrics}  \label{sec:metrics} 

Following prior work ~\cite{qian2023cross,liu2021smart,liu2021combining}, we use three common evaluation metrics to assess the performance of the SCVD methods, which are precision, recall, and F1 score. Given true positives (TP), false positives (FP), true negatives (TN), and false negatives (FN) of a classification model, precision, recall, and F1 score are defined below.

\textbf{Precision}: The proportion of correctly predicted positive instances among all instances predicted as positive. It is calculated as 
\begin{math}
	P = \frac{TP}{TP + FP}
\end{math}.

\textbf{Recall}: The proportion of correctly predicted positive instances among all actual positive instances. It is calculated as 
\begin{math}
	R = \frac{TP}{TP + FN}
\end{math}.

\textbf{F1-score}: The harmonic mean of precision and recall, which combines both metrics to provide a single measure of classification performance. It is calculated as 
\begin{math}
	F1 = 2 \times \frac{precision \times recall}{precision + recall}
\end{math}.

\subsection{Implementation Details} \label{sec:details}
For the hyperparameters of the experiment, the dimensionality of the word embeddings is set to 512. The Transformer model utilizes a six-layer multi-head attention layer with eight attention heads. During training, the learning rate is initialized to 1 x 10$^{-5}$ and is optimized using the AdamW optimizer ~\cite{loshchilov2018fixing}. The batch size is fixed at eight. The total losses in the CL stage include contrastive loss $\text{Loss}_{CL}$ and MLM loss $\text{Loss}_{MLM}$, where we set the weights for them, $\lambda_{CL}$ and $\lambda_{MLM}$, to be 1.0 and 0.1 respectively. For all the baselines, we utilize the complete code from their provided open-source libraries and adhere to the configurations specified in their research. During the CL stage, we set the number of training epochs to 100. In the vulnerability detection stage, we perform training for 20 epochs to fine-tune the model and output the result of the last epoch. The aforementioned procedure is iterated five times and the average value is chosen as the conclusive outcome of this study. 

The experiments are conducted using hardware resources that included 2 Nvidia RTX 3090 GPUs with 48GB video memory. These GPUs are utilized in parallel for training. For the software environment, we employ Ubuntu 20.04 LTS as the operating system. 

\begin{table*}[]
	\caption{The results of the ablation test.}
	\resizebox{.8\linewidth}{!}{
		\renewcommand{\arraystretch}{1.1}
		\begin{tabular}{l|rrr|rrr|rrr|rrr}
			\toprule
			\multirow{4}{*}{\textbf{Methods}} & \multicolumn{3}{c|}{\multirow{2}{*}{\textbf{RE}}} & \multicolumn{3}{c|}{\multirow{2}{*}{\textbf{TD}}} & \multicolumn{3}{c|}{\multirow{2}{*}{\textbf{IO}}} & \multicolumn{3}{c}{\multirow{2}{*}{\textbf{Average}}} \\
			& \multicolumn{3}{c|}{} & \multicolumn{3}{c|}{} & \multicolumn{3}{c|}{} & \multicolumn{3}{c}{} \\ \cline{2-13} 
			& \multirow{2}{*}{\textbf{R(\%)}} & \multirow{2}{*}{\textbf{P(\%)}} & \multirow{2}{*}{\textbf{F(\%)}} & \multirow{2}{*}{\textbf{R(\%)}} & \multirow{2}{*}{\textbf{P(\%)}} & \multirow{2}{*}{\textbf{F(\%)}} & \multirow{2}{*}{\textbf{R(\%)}} & \multirow{2}{*}{\textbf{P(\%)}} & \multirow{2}{*}{\textbf{F(\%)}} & \multirow{2}{*}{\textbf{R(\%)}} & \multirow{2}{*}{\textbf{P(\%)}} & \multirow{2}{*}{\textbf{F(\%)}} \\
			&  &  &  &  &  &  &  &  &  &  &  &  \\ \midrule
			\textbf{Clear-MVN} &75.88 &91.86 &83.11 &90.86 &74.68 &81.03 &82.54 &76.47 &79.39 &83.09 &81.00 &81.18 \\ 
			\textbf{Clear-MVV} &90.03 &81.76 &85.44 &87.30 &83.96 &85.60 &84.58 &81.94 &83.24 &87.30 &82.55 &84.76 \\
			\textbf{Clear-RMLM} &91.30 &90.23 &90.77 &93.04 &90.67 &91.85 &84.44 &88.37 &86.36 &89.59 &89.76 &89.66 \\ 
			\textbf{Clear-RCL} &63.25 &81.77 &71.32 &68.48 &81.91 &71.76 &71.62 &82.97 &76.17 &67.78 &82.22 &73.08 \\
			\textbf{Clear} &\textbf{96.43} &\textbf{96.81} &\textbf{96.62} &\textbf{98.41} &\textbf{94.30} &\textbf{96.31} &\textbf{91.48} &\textbf{89.81} &\textbf{90.64} &\textbf{95.44} &\textbf{93.64} &\textbf{94.52} \\ \bottomrule
	\end{tabular}}
	\label{tab:rq2}
\end{table*}

\section{Results}

\subsection{RQ1: Effectiveness of the Clear}

Table~\ref{tab:rq1} shows the performance evaluation of 13 SCVD methods, focusing on three prevalent and critical vulnerabilities: RE, TD, and IO. We use bold font to represent the best result of all compared approaches for each type of vulnerability. It is important to note that some vulnerability detection tools (such as Osiris, Oyente, and Securify) fail to identify all three vulnerability types. Consequently, the table includes their results only for the corresponding vulnerabilities, and we do not report an average value for these tools.

The initial comparison involves Clear and seven rule-based techniques: sFuzz, Smartcheck, Osiris, Oyente, Mythril, Securify, and Slither. The performance of these methods is presented in lines 1 to 7 of the table. We observe that Clear exhibits a substantial performance improvement over existing rule-based vulnerability detection tools across all three vulnerability types. Specifically, compared with Slither, which is the state-of-the-art tool for RE and TD detection, Clear achieves an F1-score of 96.62\% and 96.31\% in RE and TD, respectively, representing a significant improvement of 30.62\% and 41.22\%.
For IO detection,  Clear outperforms Mythril, which is the state-of-the-art tool for IO detection, by 35.69\% in terms of F1-score, achieving 90.64\%.

Subsequently, we compare Clear with five state-of-the-art deep learning-based vulnerability detection methods, including GCN, TMP, AME, SMS, and DMT. The performance results of these methods are presented in lines 8 to 12 of Table~\ref{tab:rq1}. The experimental results show the effectiveness of Clear in detecting the three vulnerability types compared to existing deep learning-based approaches. The best-performed DMT achieves average recall, precision, and F1-score of  85.13\%, 87.28\%, and 86.14\% for the three types of vulnerability. Our Clear outperforms the DMT on all three metrics. The precision and recall of Clear achieve 93.64\% and 95.44\%, respectively, representing a significant improvement of 12.11\% and 7.29\% compared to the average values obtained by DMT, resulting in an overall F1-score of 94.52\%.

Finally, in Table \ref{tab:rq1}, line 13 reports the performance of the state-of-the-art general method, LineVul, which uses CodeBERT to detect vulnerabilities. The quantitative results suggest that simple migration of general methods to the SCVD field may not yield satisfactory results. Even the LineVul, which performs the best among the general methods, only achieves an average precision, recall and F1-score of 82.92\%, 71.56\%, and 76.57\% for the three vulnerability detection scenarios, respectively. In comparison, Clear outperforms LineVul across all three metrics, surpassing LineVul by more than 12.93\%, 33.37\%, and 23.44\%, respectively.

\begin{tcolorbox}
	\textbf{Answer to RQ1}: The proposed Clear outperforms the state-of-the-art methods across all metrics. On average, Clear achieves an F1-score of 94.52\%, showcasing a 9.73\% increase in F1-score compared to the existing best-performing method.
\end{tcolorbox}

\begin{figure}
	\centering
	\includegraphics[width=\linewidth]{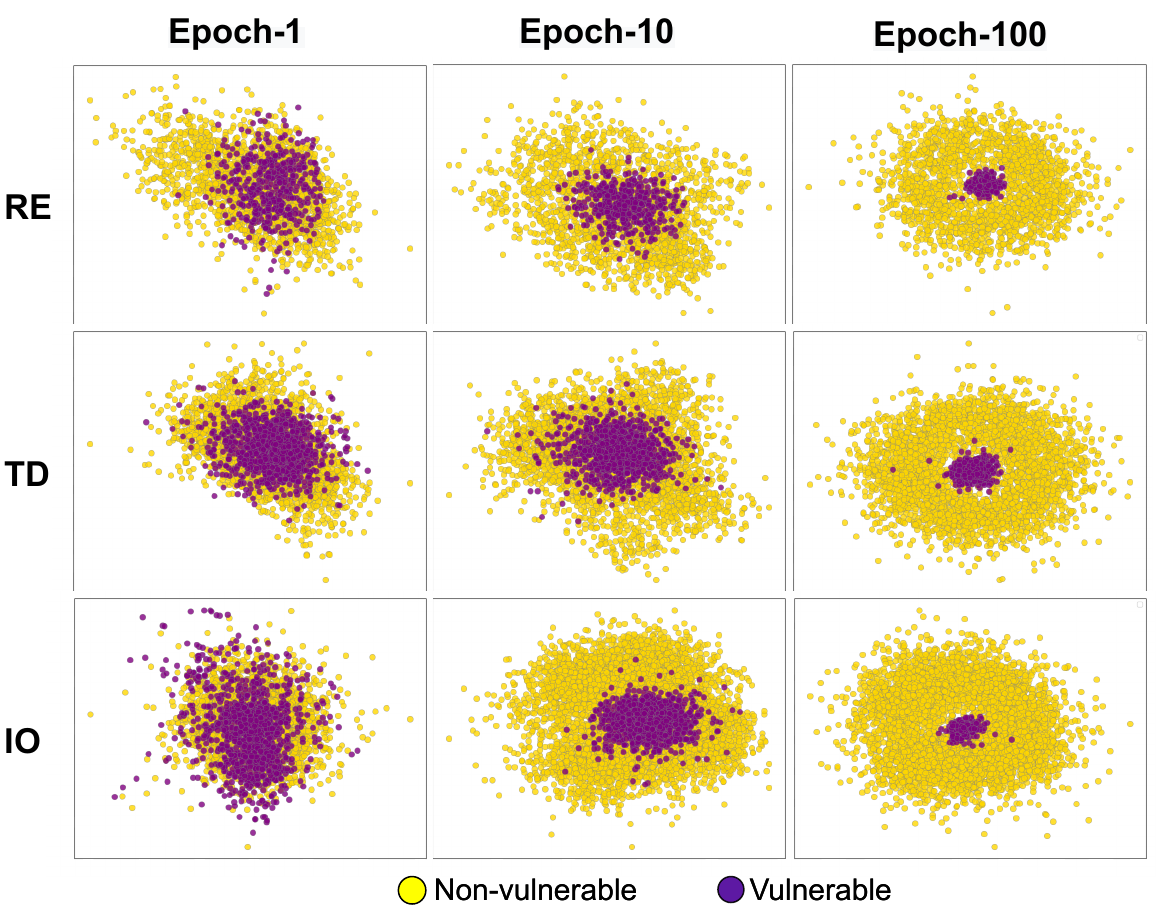}
	\caption{The feature distribution of smart contracts at different epochs during the CL stage.}
	\label{fig:rq2}
\end{figure}

\subsection{RQ2: Impact of Different Modules} \label{rq2}
To answer RQ2, we conduct comprehensive ablation tests to examine and understand the impact of different modules on Clear's overall effectiveness. In Section \ref{3-1}, we have described that Clear consists of three stages, and therefore, we have specifically designed distinct ablation tests for each of these stages. The results of all ablation tests are presented in Table \ref{tab:rq2}, in which the metrics \textbf{P}, \textbf{R}, and \textbf{F} represent precision, recall, and F1-score, respectively.

To begin with, for stage 1, we focus our data sampling strategy on two specific types of contract relationships, namely ``\textbf{V}-\textbf{V}'' and ``\textbf{V}-\textbf{N}''. We selectively mask these relationships in order to evaluate the influence of the labels generated by these two relationships on the overall effectiveness of vulnerability detection. The ``Clear-MVV'' indicates the masked ``\textbf{V}-\textbf{V}'' relationship and ``Clear-MVN'' indicates the masked ``\textbf{V}-\textbf{N}''. As shown in Table~\ref{tab:rq2}, Clear-MVN and Clear-MVV achieve 81.18\% and 84.76\% average F1-score, respectively. In comparison, Clear outperforms both of them and has an F1-score of 94.52\%. That is to say, learning only one of the contract relations within the CL stage does not yield satisfactory results. It is only by simultaneously learning both relations that we observe a significant improvement in the performance of SCVD.

\begin{table*}[h!]
	\caption{Results of RQ3.}
	\resizebox{.95\linewidth}{!}{  
		\renewcommand{\arraystretch}{1.1}
		\begin{tabular}{l|rrr|rrr|rrr|ccc}
			\toprule
			\multirow{4}{*}{\textbf{Methods}} & \multicolumn{3}{c|}{\multirow{2}{*}{\textbf{RE}}} & \multicolumn{3}{c|}{\multirow{2}{*}{\textbf{TD}}} & \multicolumn{3}{c|}{\multirow{2}{*}{\textbf{IO}}} & \multicolumn{3}{c}{\multirow{2}{*}{\textbf{Average}}} \\
			& \multicolumn{3}{c|}{} & \multicolumn{3}{c|}{} & \multicolumn{3}{c|}{} & \multicolumn{3}{c}{} \\ \cline{2-13} 
			& \multirow{2}{*}{\textbf{R(\%)}} & \multirow{2}{*}{\textbf{P(\%)}} & \multirow{2}{*}{\textbf{F(\%)}} & \multirow{2}{*}{\textbf{R(\%)}} & \multirow{2}{*}{\textbf{P(\%)}} & \multirow{2}{*}{\textbf{F(\%)}} & \multirow{2}{*}{\textbf{R(\%)}} & \multirow{2}{*}{\textbf{P(\%)}} & \multirow{2}{*}{\textbf{F(\%)}} & \multirow{2}{*}{\textbf{R(\%)}} & \multirow{2}{*}{\textbf{P(\%)}} & \multirow{2}{*}{\textbf{F(\%)}} \\
			&  &  &  &  &  &  &  &  &  &  &  &  \\ \midrule
			\textbf{RNN} & 33.60 & 37.78 & 35.56 & 46.40 & 69.05 & 55.50 & 49.46 & 58.72 & 53.70 & 43.15 & 55.18 & 48.25 \\
			\textbf{CL-RNN} & 64.29 & 59.34 & 61.71 & 76.19 & 69.06 & 72.45 & 82.08 & 69.60 & 75.33 & 74.19($\uparrow 71.94\%$) & 66.00 ($\uparrow 19.61\%$) & 69.83 ($\uparrow 50.94\%$) \\ \midrule
			\textbf{LSTM} & 35.71 & 42.66 & 38.87 & 61.11 & 63.64 & 62.35 & 55.56 & 59.39 & 57.41 & 50.79 & 55.23 & 52.88 \\
			\textbf{CL-LSTM} & 74.31 & 53.26 & 62.05 & 83.33 & 75.00 & 78.95 & 86.64 & 77.67 & 81.91 & 81.43($\uparrow 60.33\%$) & 68.64($\uparrow 24.28\%$) & 74.30 ($\uparrow 40.51\%$) \\ \midrule
			\textbf{GRU} & 50.20 & 35.88 & 41.85 & 61.11 & 63.64 & 62.35 & 67.38 & 55.95 & 61.14 & 59.56 & 51.82 & 55.11 \\
			\textbf{CL-GRU} & 80.32 & 59.69 & 68.48 & 81.60 & 78.46 & 80.00 & 89.61 & 78.86 & 83.89 & 83.84 ($\uparrow 40.77\%$) & 72.34 ($\uparrow 39.60\%$) & 77.46($\uparrow 40.56\%$) \\ \bottomrule
	\end{tabular}}
	\label{tab:rq3}
\end{table*}

Moving on to stage 2, we intentionally remove the MLM module that is integrated into the CL stage. This allows us to analyze the overall effectiveness of the CL stage without the presence of the MLM module. This particular test is referred to as ``Clear-RMLM''. We observe that the MLM module has a substantial impact on the effectiveness of the CL module. Specifically, when the MLM module is removed, there is an average decrease in precision, recall, and F1-score by 4.15\%, 6.13\%, and 5.14\% respectively. Therefore, we believe that the MLM module can enhance the performance of Clear and is an essential component.

Lastly, for stage 3, we remove the CL stage altogether and directly performed the vulnerability detection stage. This test, known as ``Clear-RCL'', enables us to evaluate the performance of vulnerability detection in the absence of the CL stage. In comparison to Clear-RCL, we observe a significant improvement in performance for all three types of vulnerability detection tasks with the addition of the CL module. The F-score increased by 35.47\% for RE, 34.21\% for TD, and 19.00\% for IO. This notable improvement can be attributed to the synergistic effects of the CL stage itself and our unique sampling strategy. Specifically, the CL module facilitates the convergence of dispersed vulnerability samples in the feature space, resulting in increased proximity among them. By utilizing our unique sampling strategy, we further reinforce the correlation among samples belonging to the same vulnerability category, thereby promoting their clustering behavior. This process enables the model to more effortlessly identify and discover potential SCVs, leading to a significant improvement in the performance of SCVD.

To substantiate our assertion, we thoroughly examine the derivation process of the sample distribution during the CL stage. In particular, we analyze the evolution of the output of the CL stage (denoted as $\bm{v}$ in Eq. \ref{ClOutput}) at each epoch and employ principal component analysis ~\cite{roweis1997algorithms} to project each output onto a two-dimensional space. Subsequently, these outputs are visualized as scatter plots and displayed in Figure \ref{fig:rq2}, where the horizontal and vertical axes represent linear combinations of the vectors $\mathbf{v}$ obtained through PCA. Each point denotes a contract sample, with purple indicating vulnerability samples and yellow representing non-vulnerability samples. The figure clearly depicts the progression of smart contract sample distribution throughout the CL stage and yields the following finding. First, during the training process of the CL module, the samples of vulnerability contracts exhibit a tendency to cluster together, while being distinctly separated from non-vulnerability samples, indicating a clear distinction between the two groups. Second, this distribution enhances the ability to differentiate and detect SCVs. Clear exhibits a higher proficiency in recognizing this particular cluster and accurately classifying contracts within its proximity as vulnerable. This leads to an improved capability for identifying SCVs.

\begin{tcolorbox}
	\textbf{Answer to RQ2}: The two types of relations in contracts are indispensable in the CL stage. The MLM module can enhance the performance of Clear. The CL module facilitates the aggregation of dispersed vulnerability samples in the feature space, leading to a significant performance improvement in the tasks of SCVD. 
\end{tcolorbox}

\subsection{RQ3: Effectiveness of the CL Module}
To further investigate the contribution of the CL module to other deep learning models in SCVD, we have selected a set of traditional deep learning models, including RNN, LSTM, and GRU, to replace the Transformer model used in Clear. These models, collectively referred to as “CL-Mode,” have been specifically chosen because Clear is designed to process code sequences directly, while the deep learning-based methods in the baselines are constructed on graph structures. Therefore, our Clear is incompatible with these graph-based methods. Additionally, we present the performance results of traditional deep learning models to facilitate comparisons and analysis. 

The statistical results of these experiments are presented in Table \ref{tab:rq3} and provide the following observations. All models exhibit notable enhancements in performance when embedded with the CL module. In terms of F1-score, CL-RNN, CL-LSTM, and CL-GRU improved 44.71\% (from 48.25\% to 69.83\%), 40.52\% (from 52.88\% to 74.30\%) and 40.54\% (from 55.11\% to 77.46\%) over the original model, respectively, on the average value of three vulnerabilities. 

\begin{tcolorbox}
	\textbf{Answer to RQ3}: The empirical evidences suggest the potential of combining traditional deep learning models with the CL module for SCVD, which facilitates the model in acquiring fine-grained feature information and enhancing the performance of SCVD.
\end{tcolorbox}

\section{Related Work} \label{RelatedWork}

\subsection{Smart Contract Vulnerability Detection}
SCVD is an important research problem in blockchain security, and numerous scholarly works have been dedicated to exploring it. The initial approaches to detecting SCVs involved static analysis and dynamic execution techniques based on some predefined rules or patterns. For example,
Oyente ~\cite{luu2016making} was one of the early SCV detection methods that utilized symbolic execution. It focused on detecting vulnerabilities by analyzing the contract's control flow graph based on symbolic execution. Securify ~\cite{tsankov2018securify} examined the contract's dependency graph and extracted detailed semantic information from the code to identify compliance and security vulnerabilities.
Mythril ~\cite{mueller2017framework} was a static analysis tool that employed concept analysis, taint analysis, and control flow verification to detect common SCVs. TeEther ~\cite{krupp2018teether} analyzed the contract bytecode and searched for critical execution paths to identify SCVs. 
Slither ~\cite{feist2019slither} was a static analysis framework that converted smart contract source code into an intermediate representation called SlithIR. It utilized this representation to detect SCVs. 
Osiris ~\cite{torres2018osiris} combined symbolic execution and taint analysis techniques to detect integer errors in smart contracts. 
SmartCheck ~\cite{tikhomirov2018smartcheck}, another static program analysis tool, converted Solidity source code into XML and checked for vulnerabilities based on predefined XPath patterns. 
sFuzz ~\cite{nguyen2020sfuzz} employed a branch distance-driven fuzzing technique to identify vulnerabilities. 
SMARTIAN ~\cite{choi2021smartian} was a fuzzier, which utilized lightweight dynamic data-flow analysis to guide fuzzing by collecting feedback based on data flow.

With the advancement of deep learning, there has been a rise in research approaches that harness automated deep learning methods for smart contract vulnerability detection. For example,
SaferSC ~\cite{tann2018towards} was the first vulnerability detection method to utilize deep learning. It employed a Long Short-Term Memory (LSTM) network to construct a sequence model of Ethereum opcode, providing a comprehensive representation to detect vulnerabilities.
More recent deep learning research in this field emphasizes the use of graph structures. 
DR-GCN ~\cite{zhuang2021smart} transformed smart contract source code into a contract graph with high semantic representation and employed a Graph Convolutional Neural Network (GCN) to construct a vulnerability detection model.
TMP ~\cite{zhuang2021smart} extended the approach of DR-GCN by converting critical functions and variables into core nodes with rich semantic information within the contract graph. It also incorporated temporal information on edges.
CGE ~\cite{liu2021combining} built upon TMP by further incorporating expert mode information, integrating the contract graph information with expert knowledge.
AME ~\cite{liu2021smart} aimed to combine deep learning and expert mode in an interpretable manner, building upon the CGE approach.
DMT ~\cite{qian2023cross} proposed a single-modality student network and a cross-modality mutual learning framework to enhance smart contract vulnerability detection on bytecode.
However, it is worth noting that all of the methods mentioned above primarily focus on detecting vulnerabilities by learning the semantic knowledge of current input contracts and ignoring the correlation between contracts. Indeed, the inter-contract correlation plays a critical role in understanding the overall security of smart contract ecosystems. Our method successfully improves the performance of SCVD by incorporating correlation information.

\subsection{General Vulnerability Detection}

Moreover, we also investigate some traditional vulnerability detection techniques that are used to detect vulnerabilities in JAVA, C, and C++ programming languages. The Devign model, proposed by Zhou et al. ~\cite{zhou2019devign}, is a generalized graph neural network-based approach for detecting program vulnerabilities. Its effectiveness was validated through experiments conducted on four different large-scale open-source C projects. Li et al. ~\cite{li2021vulnerability} introduced IVDetect, an interpretable vulnerability detector that leverages deep learning techniques to model program dependency graphs for the purpose of detecting vulnerabilities. Fu et al. ~\cite{fu2022linevul} proposed LineVul, a Transformer-based approach for detecting vulnerabilities of the C/C++ program at the line level. By utilizing pre-trained models to learn fine-grained code semantic information, LineVul has proven to be the most effective and highest-performing method available. The aforementioned methods, however, fail to yield satisfactory outcomes in the domain of smart contract vulnerability detection, even with LineVul being considered as the most effective approach, there exists a discernible performance disparity when compared to our method.

\subsection{Contrastive Learning}

The CL was initially developed as an unsupervised learning technique, aiming to learn representations by uncovering underlying similarities and dissimilarities among samples.

In the field of computer vision, several unsupervised learning methods have proposed CL techniques. Notably, InstDisc \cite{wu2018unsupervised} was an unsupervised method widely used in computer vision for learning the representation of data. Its goal was to map samples from the same class to similar representation spaces, while samples from different classes were mapped to different representation spaces. InvaSpread \cite{ye2019unsupervised} proposed an end-to-end learning mechanism that could perform positive and negative sample comparisons within the same mini-batch. MoCo \cite{he2020momentum} introduced a contrastive learning method based on a dynamic dictionary and dynamic negative sample queue, which improved the quality of feature representation by constructing a large dynamic dictionary to extend the positive sample set. SimCLR \cite{chen2020simple} was a simple framework for contrast learning representations. SimCLR learned image representation by maximizing the similarity between different views of the same image and achieved significant performance improvements on tasks such as image classification, object detection, and semantic segmentation. SwAV \cite{zhu2020swav} learned image representation by introducing the idea of clustering, assigning samples to different cluster clusters, and achieved impressive results on tasks such as unsupervised image segmentation, object detection, and image classification. SimSiam \cite{chen2021exploring} proposed a simple framework for self-supervised learning to learn the representation of images or features. The core idea was to learn the representation of features by minimizing the Euclidean distance between different views of the same sample using an autoencoder. Its advantages lay in its simplicity and efficiency, without the need to use complex contrast loss functions or negative sample mining strategies.

In the NLP field, ConSERT \cite{qiu2021contrastive} utilized various data augmentation techniques to construct positive sample pairs, such as cutoff, shuffle, adversarial learning, and dropout. SimCSE \cite{gao2021simcse} used a simple "dropout twice" technique to construct positive sample pairs for CL, achieving a new state-of-the-art performance in unsupervised semantic similarity tasks. ESimCSE \cite{wu2021esimcse} later introduced a momentum CL method to construct negative sample pairs. The R-Drop method is similar to SimCSE, applying the "dropout twice" technique to supervised tasks.

Notably, we employ CL for the first time in the SCVD domain and utilize correlation labels to guide the training of the CL model, which play a crucial role in fitting correlation features by the model and effectively enhance the performance of SCVD.

\section{Threats of Validity}
\textbf{Threats to external validity} arise from the datasets and studied vulnerabilities. To minimize the former threat, we utilize the largest publicly available vulnerability dataset that consists of smart contracts labeled as either vulnerable or non-vulnerable. Additionally, we focus on evaluating the studied vulnerability detection methods on the three most severe and common vulnerabilities, further enhancing the external validity of our research. 

\textbf{Threats to internal validity} stem from the implementation of Clear and the compared vulnerability detection methods. To address these threats, we implement Clear using the PyTorch framework and leverage established third-party libraries. Moreover, we utilize the reproducible package of the compared methods, ensuring a fair and standardized comparison.

\textbf{Threats to construct validity} arise from the metrics used to measure the performance of the studied vulnerability detection methods. To mitigate these threats, we employ widely accepted evaluation metrics, such as precision, recall, and F1-score. These metrics provide a comprehensive assessment of the classification performance, ensuring the construct validity of our research. Besides, hardware devices significantly influence the speed of detection as well as threaten structural validity. The threat is addressed by conducting all experiments in this paper on the same device, resulting in traditional detection tools taking approximately 20 to 60 seconds to detect a smart contract, while deep-learning-based detection methods require less than 1 second for the same task.

\section{Conclusion}
With the rapid development of blockchain technology, transactions based on smart contracts have become increasingly frequent and security has become even more critical. However, the SCVs have emerged as one of the top threats to secure transactions. While numerous methods have been successful in mitigating this threat, the discriminative power of existing methods on SCVs still has a lot of room for improvement. Because they fail to explore fine-grained information from vulnerability labels and take into account correlations among contracts. To address these issues, we propose the Clear model, which leverages the CL model to effectively capture inter-contract correlations. By introducing correlation labels, the model can learn fine-grained correlation information. To validate the effectiveness of our Clear, we conduct extensive experiments on a dataset consisting of over 40,000 real-world smart contracts. These contracts are evaluated against state-of-the-art detection methods and compared to Clear for performance. The results demonstrate that our proposed Clear outperforms all detection methods, thereby improving the overall effectiveness of SCVD.

\begin{acks}
This work was supported by National Natural Science Foundation of China under Grant Nos. 62372005 and 62232001.
\end{acks}

\bibliographystyle{ACM-Reference-Format}
\bibliography{bibfile}
\appendix

\end{document}